\documentclass[10pt,twocolumn,preprintnumbers,amsmath,amssymb,nofootinbib
,superscriptaddress, floatfix]{revtex4-2}

\usepackage{graphicx}
\usepackage{aas_macros}
\usepackage{dcolumn}
\usepackage{bm}
\usepackage[utf8]{inputenc}
\usepackage{color}
\usepackage[colorlinks,linkcolor=blue,citecolor=red,urlcolor=blue ]{hyperref}
\usepackage[dvipsnames]{xcolor}
\usepackage{mathptmx}
\usepackage{xspace}
\usepackage{booktabs}
\usepackage{multirow}

\newcommand{\nv}{\hat{\bf n}}
\newcommand{\Ow}{\Omega_{\rm GW}}
\newcommand{\bOw}{\langle b\dot{\Omega}_{\rm GW}\rangle}
\newcommand{\lgw}{{\cal L}_{\rm GW}}
\newcommand{\tmpz}{2MPZ\xspace}
\newcommand{\wisc}{WI$\times$SC\xspace}
\newcommand{\dels}{DECaLS\xspace}
\newcommand{\quaia}{Quaia\xspace}
\makeatletter
\newcommand*{\rom}[1]{\expandafter\@slowromancap\romannumeral #1@}
\makeatother

\begin{document}

\title{Tomographic constraints on the production rate of gravitational waves from astrophysical sources}

\author{David Alonso}
\affiliation{Department of Physics, University of Oxford, Denys Wilkinson Building, Keble Road, Oxford OX1 3RH, United Kingdom}
\author{Mehraveh Nikjoo}
\affiliation{Institute of Theoretical Physics and Astrophysics, University of Gda\'nsk, 80-308 Gda\'nsk, Poland}
\author{Arianna I. Renzini}
\affiliation{Dipartimento di Fisica ``G. Occhialini'', Universit\'a degli Studi di Milano-Bicocca, Piazza della Scienza 3, 20126 Milano, Italy}
\affiliation{INFN, Sezione di Milano-Bicocca, Piazza della Scienza 3, 20126 Milano, Italy}
\author{Emilio Bellini}
\affiliation{SISSA, International School for Advanced Studies, Via Bonomea 265, 34136 Trieste, Italy}
\affiliation{IFPU, Institute for Fundamental Physics of the Universe, via Beirut 2, 34151 Trieste, Italy}
\affiliation{INFN, National Institute for Nuclear Physics, Via Valerio 2, I-34127 Trieste, Italy}
\author{Pedro G. Ferreira}
\affiliation{Department of Physics, University of Oxford, Denys Wilkinson Building, Keble Road, Oxford OX1 3RH, United Kingdom}
\date{\today}

\begin{abstract}
  Using an optimal quadratic estimator, we measure the large-scale cross-correlation between maps of the stochastic gravitational-wave intensity, constructed from the first three LIGO-Virgo observing runs, and a suite of tomographic samples of galaxies covering the redshift range $z\lesssim 2$. We do not detect any statistically significant cross-correlation, but the tomographic nature of the data allows us to place constraints on the (bias-weighted) production rate density of gravitational waves by astrophysical sources as a function of cosmic time. Our constraints range from $\bOw<3.0\times10^{-9}\,{\rm Gyr}^{-1}$ at $z\sim0.06$ to  $\bOw<2.7\times10^{-7}\,{\rm Gyr}^{-1}$ at $z\sim1.5$ (95\% C.L.), assuming a frequency spectrum of the form $f^{2/3}$ (corresponding to an astrophysical background of binary mergers), and a reference frequency $f_{\rm ref}=25\,{\rm Hz}$. Although these constraints are $\sim2$ orders of magnitude higher than the expected signal, we show that a detection may be possible with future experiments.
\end{abstract}

\maketitle

\section{Introduction}\label{sec:intro}
 The detection of the stochastic gravitational-wave background (SGWB) represents a new challenge for astrophysics and cosmology. Measurements of the SGWB will, in principle,  unveil a wealth of information about the early Universe~\cite{cosmo_SGWB_review}, the large scale structure of the cosmos~\cite{Renzini:2022alw}, and astrophysical phenomena~\cite{Regimbau_astro_SGWB} that remain hidden from conventional electromagnetic observations. This background, formed by the superposition of countless unresolved gravitational-wave (GW) sources, constitutes a complex and persistent signal that permeates the fabric of spacetime. One hopes that a judicious analysis of a putative SGWB will allow us to disentangle the different processes at play and will lead to new insights into a range of phenomena.

There has been substantial progress in GW astronomy through the efforts of observatories like the Laser Interferometer Gravitational-wave Observatory (LIGO,~\cite{LIGO}), Virgo~\cite{Virgo}, and KAGRA~\cite{KAGRA}, which have significantly improved the ability to detect individual gravitational-wave events since the first direct detection in 2016~\cite{LIGOScientific:2016aoc}. Progress has also been made towards constraining the mean, or monopole, of the gravitational-wave background through dedicated GW searches which employ a stochastic signal model, as seen in works~\cite{iso-O2, iso-O3, Renzini:2023qtj} by the LIGO-Virgo-KAGRA Collaboration (LVK). %
As of the most recent completed LVK observing run, O3, the projected astrophysical GWB is roughly an order of magnitude smaller than the most competitive upper limits placed by the collaboration~\cite{iso-O3}; the expectation is that the signal will start becoming detectable once ``Design A+'' sensitivity is reached~\cite{sensitivity_AplusDesign}. %
A very recent encouraging stochastic success story is that of pulsar timing arrays (Nanograv~\cite{NANOGrav:2023gor},
EPTA~\cite{EPTA:2023fyk}, Parkes~\cite{PPTA}, CPTA~\cite{CPTA}), which announced strong evidence for a GWB at nanohertz frequency scales, marking the beginning of the stochastic GW detection era. %
The challenge of isolating the faint, anisotropic, SGWB from the more prominent signals remains formidable and involves sophisticated data analysis techniques that can tease the signal out from a myriad of systematic -- instrumental or astrophysical -- effects. 

There are two dominant sources of noise that hinder attempts at measuring the SGWB. The first is the detector noise of the experimental setup. In~\cite{2005.03001} the authors showed that such noise would imprint spurious, stochastic, spatial structure on any attempts to reconstruct maps of the SGWB anisotropies. %
Furthermore, in both ground-based and space-based observatories (for example, the Laser Interferometer Space Antenna~\cite{LISA}), the spatial power spectrum of the noise would likely swamp any anisotropy in a cosmological signal from the early or late Universe. %
The second source of noise is the fact that the astrophysical sources of gravitational waves are, in effect, discrete, leading to what has been dubbed a "popcorn" signal -- a white noise akin to the "shot" noise found in cosmological surveys (see e.g.~\cite{Jenkins:2019uzp}). Such noise can completely dominate any large scale features of the SGWB, and needs to be appropriately mitigated (see e.g.~\cite{Kouvatsos:2023bgd} for a proposed de-biasing approach).

One way to mitigate both of these contaminants is to use a template to enhance the signal. In the case of the SGWB an obvious template is some tracer of the large scale structure (LSS) of the Universe which is hosting the sources of gravitational waves. By cross-correlating a tracer of the LSS with a map of the SGWB, it should be possible to enhance or "pull out" the gravitational waves that were generated by astrophysical sources correlated with this LSS \citep{2020PhRvD.102b3002A}. Furthermore, by studying cross-correlations of any given map with galaxies at different redshifts, one can place constraints on the intensity of the physical processes that generate this map as a function of cosmic time. This technique, commonly known as ``tomography'', has been employed in the literature to reconstruct multiple astrophysical and cosmological observables \citep{1808.03294,1810.00885,1909.09102,2006.14650,2105.12108,2206.15394,2307.14881}.  Given that we are in an age in which LSS data is greatly increasing in quality and quantity, it would seem that this is a sensible approach to take to place constraints on the sources of the SGWB.

There have been attempts at cross-correlating maps of the SGWB with LSS. In the most recent analysis ~\cite{yang2023measurement}, the authors used the number counts from the Sloan Digital Sky Survey  (SDSS) to place constraints on gravitational-wave emission at a median redshift, $z=0.39$. Furthermore, they divided the SGWB map in $10$ Hz bins in an attempt to extract spectral information. Their analysis resulted in constraints on the gravitational-wave emission, over a range of scales, of around $\sim {\rm few} \, \times 10^{-32}\, {\rm erg}\, {\rm cm}^{-3}\, {\rm s}^{-1/3}$.

In this paper, we present improved constraints on the bias-weighted gravitational-wave emission using SGWB maps from the latest LVK dataset~\cite{2211.10010}. Our approach differs from that in ~\cite{yang2023measurement} in several ways: we use an optimal quadratic minimum variance estimator (QMVE) to extract the large-scale cross-correlation, we consider 4 different tomographic samples (\tmpz, \wisc, \dels, and \quaia) spanning the redshift range $z\lesssim 2$, and we transform our measurements into constraints on a physical parameter, $\bOw$, the bias-weighted average production rate of gravitational waves by astrophysical sources.

The structure of the paper is as follows. In Section \ref{sec:theory} we outline the theoretical model for the energy density of gravitational waves in relation to bias-weighted mean emissivity and the angular cross-power spectrum between galaxies and maps of the astrophysical GWB. Section \ref{sec:data} describes the SGWB maps and galaxy samples used in our analysis. Section \ref{sec:meth} details the methods used to estimate the power spectrum as well as the likelihood used to extract parameter constraints from the data. Finally, we discuss the obtained results for the measurement of the power spectrum and the achieved tomographic constraints to attain an upper bound on the bias-weighted production rate of GW density in section \ref{sec:res}, as well as presenting forecasts for similar measurements carried out by future experiments. We summarise and conclude in Section \ref{sec:conc}.

\section{Astrophysical SGWB modelling and cross-correlations}\label{sec:theory}
  \subsection{Astrophysical SGWB}\label{ssec:theory.sgwb}
    The stochastic GWB is commonly quantified in terms of the fractional energy density in gravitational waves averaged per unit solid angle, $\Ow$, defined as \cite{1908.00546}:
    \begin{equation}
      \Ow\equiv\frac{1}{\rho_{c,0}}\frac{dE_o}{d\log f_o\,dV_o\,d\Omega_o},
    \end{equation}
    where $dE_o$, $df_o$, $dV_o$, and $d\Omega_o$ are intervals of GW energy, frequency, volume, and solid angle, measured in the observer's frame (hence the $_o$ subscript), and $\rho_{c,0}\equiv 3H_0^2/8\pi G$ is the critical density today. 

    Our derivation of the SGWB anisotropies from astrophysical sources below is inspired by the extensive existing literature in this topic (e.g. \cite{1704.06184,1711.11345,1803.03236,2106.02591,2110.15059}).

    Consider now the contribution to the value of $\Ow$ measured in a given direction $\nv$ by sources distributed along the line of sight. The relation between $\Ow$ and the gravitational-wave emissivity $j_{\rm GW}$ is\footnote{We use natural units, with $c=1$, throughout.}
    \begin{equation}\label{eq:Ow1}
      \Ow(\nv) = \int d\chi\frac{f_o}{4\pi\,\rho_{c,0}(1+z)^4}j_{\rm GW}(\chi\nv,z),
    \end{equation}
    where $\chi$ is the comoving radial distance to redshift $z$, $j_{\rm GW}$ is the energy output per unit time and volume in the emitter's frame (subscripted as $_e$)
    \begin{equation}
      j_{\rm GW}\equiv\frac{dE_e}{dt_e\,df_e\,dV_e},
    \end{equation}
    and the factor $(1+z)^{-4}$ accounts for the redshifting of energy and clock rates at the observer and emitter, as well as the relation between transverse areas (Etherington relation). Note that, in Eq. \ref{eq:Ow1}, we have omitted the implicit dependence of $\Omega_{\rm GW}$ on frequency $f_o$. This is because, under the assumption of a factorisable spatial and frequency dependence (as described in Section \ref{ssec:data.sgwb}), we will consider cross-correlations against a single GWB map at an effective frequency $f_{\rm ref}$ assuming a particular spectrum.

    Assuming the SGWB to be dominated by astrophysical sources, the GW emissivity can be written in terms of the GW luminosity function of such sources $dn/d\lgw$:
    \begin{equation}
      j_{\rm GW}=\int d\lgw\,\frac{dn}{d\lgw},
    \end{equation}
    where $\lgw\equiv dE_e/(dt_e\,df_e)$ is the specific GW luminosity. Expanding the luminosity function in terms of its global average and the spatial fluctuations in the density of sources, $\delta_s$, due to the presence of matter inhomogeneities, we can write the anisotropies (i.e. angular fluctuations in $\Ow$ with respect to the sky average) as
    \begin{align}\nonumber
      \Delta\Ow(\nv)&=\int d\chi\,\frac{f_o}{4\pi\,c\,\rho_{c,0}(1+z)^4}\\
      &\hspace{20pt}\int d\lgw\,\lgw\,\frac{d\bar{n}}{d\lgw}\,\delta_s(\chi\nv,z,\lgw).
    \end{align}
    On sufficiently large scales ($k\lesssim0.1\,{\rm Mpc}^{-1}$, appropriate to the measurements analysed in this work), a simple linear bias relation may be assumed to connect the source overdensity $\delta_s$ and the underlying matter overdensity $\delta$. In this limit, we can then write
    \begin{align}\nonumber
      \Delta\Ow(\nv)&=\int d\chi\,\frac{f_o\langle bj_{\rm GW}(z)\rangle}{4\pi\,c\,\rho_{c,0}(1+z)^4}\delta(\chi\nv,z),
    \end{align}
    where the bias-weighted mean emissivity is
    \begin{equation}
      \langle bj_{\rm GW}(z)\rangle\equiv \int d\lgw\,\lgw\,\frac{d\bar{n}}{d\lgw}\,b(\lgw,z),
    \end{equation}
    and $b(\lgw,z)$ is the linear bias of sources with GW luminosity $\lgw$. Finally, we can write $j_{\rm GW}$ in terms of the more suggestive fractional energy density rate $\dot{\Omega}_{\rm GW}$:
    \begin{equation}
      \dot{\Omega}_{\rm GW}\equiv\frac{1}{\rho_{c,e}}\frac{dE_e}{dt_e\,d\log f_e\,dV_e}=\frac{f_e}{\rho_{c,e}}j_{\rm GW}.
    \end{equation}
    We thus obtain
    \begin{equation}
      \Delta\Omega_{\rm GW}(\nv)=\int d\chi\,q_{\rm GW}(\chi)\,\bOw\,\delta(\chi\nv,z),
    \end{equation}
    where the radial kernel $q_{\rm GW}(\chi)$ is
    \begin{equation}
      q_{\rm GW}(\chi)\equiv\frac{E^2(z)}{4\pi(1+z)^5},
    \end{equation}
    with $E(z)\equiv H(z)/H_0$ the ratio of the expansion rate at redshift $z$ to its value today.

  \subsection{Cross-correlations and tomography}\label{ssec:theory.xcorr}
    Consider now the projected overdensity $\Delta_g$ for a sample of galaxies with redshift distribution $p_g(z)$. Assuming again a linear bias relation with galaxy bias $b_g$, we can write
    \begin{equation}
      \Delta_g(\nv)=\int\,d\chi\,q_g(\chi)\,b_g\,\delta(\chi\nv,z),
    \end{equation}
    where the radial kernel is proportional to the redshift distribution $q_g(\chi)\equiv H(z)\,p_g(z)$.

    With this notation, the angular cross-power spectrum between galaxies and maps of the astrophysical GWB is
    \begin{widetext}
    \begin{equation}
      C_\ell^{g{\rm GW}}=\frac{2}{\pi}\int dk\,d\chi_1\,d\chi_2\,b_g\,\bOw\,q_g(\chi_1)\,j_\ell(k\chi_1)\,q_{\rm GW}(\chi_2)\,j_\ell(k\chi_2)\,P(k,\bar{z}),
    \end{equation}
    \end{widetext}
    where $j_\ell(x)$ is the spherical Bessel function of order $\ell$, $P(k,z)$ is the matter power spectrum, and we have used the shorthand\footnote{This approximation effectively assumes perfect correlation between the overdensity field at different redshifts. This is exact in the linear regime, which is appropriate for the large scales used in this analysis.}: $P(k,\bar{z})\equiv\sqrt{P(k,z_1)P(k,z_2)}$. A similar expression would hold for the galaxy auto-correlation $C_\ell^{gg}$, simply replacing $\bOw\,q_{\rm GW}$ with $b_gq_g$.

    Assuming a reasonably compact sample of galaxies, the galaxy kernel $q_g$ is sharply peaked at the mean redshift of the sample $z_g$. Since $\bOw$ is approximately constant over the narrow support of $q_g$, we can thus simply evaluate it at $z_g$ and pull it outside of the integral in the previous equation. Doing so, we can write
    \begin{equation}\label{eq:cl_tomo}
      C_\ell^{gg}=b_g^2\,T_\ell^{gg},\hspace{12pt}
      C_\ell^{g{\rm GW}}=b_g\bOw\,T_\ell^{g{\rm GW}},
    \end{equation}
    where the template power spectra $T^{gg}_\ell$ and $T^{g{\rm GW}}_\ell$ depend only on cosmological parameters and the redshift distribution of the galaxy sample, and are given by:
    \begin{widetext}
    \begin{align}\label{eq:tgg}
      &T_\ell^{gg}=\frac{2}{\pi}\int dk\,d\chi_1\,d\chi_2\,\,q_g(\chi_1)\,j_\ell(k\chi_1)\,q_g(\chi_2)\,j_\ell(k\chi_2)\,P(k,\bar{z})\simeq\int \frac{d\chi}{\chi^2}q_g^2(\chi)\,P\left(k=\frac{\ell+1/2}{\chi},z\right),\\\label{eq:tgw}
      &T_\ell^{g{\rm GW}}=\frac{2}{\pi}\int dk\,d\chi_1\,d\chi_2\,\,q_g(\chi_1)\,j_\ell(k\chi_1)\,q_{\rm GW}(\chi_2)\,j_\ell(k\chi_2)\,P(k,\bar{z})\simeq\int \frac{d\chi}{\chi^2}q_g(\chi)q_{\rm GW}(\chi)\,P\left(k=\frac{\ell+1/2}{\chi},z\right).
    \end{align}
    \end{widetext}
    The simpler expressions on the right-hand side of the equations above hold in the Limber approximation. Since our analysis relies on the use of large angular scales ($\ell\lesssim 20$ in the case of $C_\ell^{g{\rm GW}}$), we will not make use of the Limber expressions, and we only provide them for completeness.

    As is evident from Eq. \ref{eq:cl_tomo}, the cross-correlation between $\Delta\Omega_{\rm GW}$ and the galaxy overdensity can be used, in combination with the galaxy auto-correlation, to simultaneously measure the bias-weighted fractional GW production rate $\bOw$ and the galaxy bias. Moreover, since the templates $T_\ell^{xy}$ depend only on cosmological parameters (which we assume to be known), $(\bOw,b_g)$ are trivially constrained from the amplitude of the measured power spectra (see Section \ref{ssec:meth.like}). This constitutes the tomographic reconstruction formalism, which has been used in the literature to study a variety of physical quantities from cross-correlations with different tracers of the large-scale structure \citep{1808.03294,1810.00885,1909.09102,2006.14650,2105.12108,2206.15394,2307.14881}, and which follows the same principle as the ``clustering redshifts'' approach to reconstructing redshift distributions of photometric galaxy samples \citep{0805.1409,1303.4722,1609.09085}.

\section{Data}\label{sec:data}
  \subsection{SGWB maps}\label{ssec:data.sgwb}

    SGWB maps are obtained through a maximum-likelihood stochastic GW analysis analogous to those performed in~\cite{ O2_anisotropic, AR_O1, AR_O1O2, O3_SGWB_anisotropic, O3_ASAF}. Specifically, here we employ pixel map estimates of $\Ow(\nv)$ presented in~\cite{2211.10010}, obtained using the most recently publicly released LVK dataset O3. %
    We briefly summarize here assumptions and methodologies employed in~\cite{2211.10010} --  we refer the reader to the full publication for further details. 
    
    Maps of the GW sky with LVK data are obtained by averaging cross-spectrograms of the full dataset ($\mathcal{O}(1)$ year), under the assumption of Gaussian stationary noise~\cite{Renzini:2023qtj}. Each spectrum is estimated over a time chunk of duration $\tau\sim \mathcal{O}(100)s$, which is the time scale of the stationarity assumption. %
    The data are optimally compressed (or {\it folded}) to a single sidereal day by leveraging the periodicity of the {\it baseline} (i.e., detector pair) response $\gamma$ (i.e., the {\it beam})~\cite{data_folding_fast, data_folding_veryfast}. %
    The result of this compression procedure is a single cross-spectrum estimate $\hat{C}$ with associated variance $\sigma_{\hat{C}}$ for each baseline, which can be averaged across baselines~\cite{O3_SGWB_anisotropic} to achieve a combined estimate. 
    
    The SGWB signal is assumed to be Gaussian and unpolarized, with a power-law spectral shape $\propto f^\alpha$~\cite{Renzini:2022alw}. In the case of an astrophysical signal arising from compact binary inspiral, the spectral index is assumed to be $\alpha = 2/3$, while for signals of cosmological origin the scale-invariant  $\alpha=0$ is used. %
    In the analysis presented here, we show results for both these spectral assumptions and also include $\alpha=3$, in line with~\cite{O3_SGWB_anisotropic, 2211.10010}, which may be considered a ``best-fit'' spectral index which traces the detector noise spectrum, leading to the most sensitive result (or most stringent upper limit). %
    Assuming the spectral and directional dependence of $\Ow(\nv)$ are independent, the signal anisotropy is projected into the datasetream through the operator $K$, which takes into account the coupling of the spectral shape of $\Ow(\nv)$ with that of the directional response of the detector, 
    \begin{equation}
        K_{t f}(\nv) \equiv \tau \left( \frac{f}{f_{\rm ref}}\right)^{3-\alpha} \gamma(t; f, \nv). 
    \end{equation}
    The above takes into account the $f^3$ scaling between strain power spectral density and $\Ow$, and $f_{\rm ref}$ is the measurement reference frequency. %
    
    Under these assumptions, the likelihood of measuring the cross-spectrum $\hat{C}$ given the strain power spectral density map $\mathcal{P}(\nv)$ is (using short-hand notation presented in~\cite{2211.10010})
    \begin{equation} \label{eq:GWB_likelihood}
        \mathcal{L} (\hat{C} | \mathcal{P}) \propto \prod_{tf}\exp \bigg[-\frac{1}{2} (\hat{C} - K \cdot \mathcal{P})^\dag N^{-1} (\hat{C} - K \cdot \mathcal{P})) \bigg].
    \end{equation}
    Maximizing Eq.~\eqref{eq:GWB_likelihood} reduces to a minimum $\chi^2$ solution, yielding the pixel estimate of the GWB ``clean map'' 
    \begin{equation}\label{eq:clean_map}
        \hat{\mathcal{P}}_\eta = \sum_{\eta'}\Gamma^{-1}_{\eta\eta'} X_{\eta'},
    \end{equation}
    where $X$ is often referred to as the ``dirty map'' and $\Gamma$ is the Fisher information matrix (see~\cite{2211.10010} for the full expressions). %
    Note that this is a broad-band estimate of the directional strain power, quoted at the reference frequency $f_{\rm ref}$; the marginalization over frequency is necessary to boost detection statistics. %
    The associated noise covariance matrix is
    \begin{equation}\label{eq:NGW}
        {\sf N}_{\rm GW} = \Gamma_{\eta \eta'}^{-{1}}\,.
    \end{equation}

    Calculating ${\sf N}_{\rm GW}$ thus requires the inversion of the Fisher pixel--pixel matrix, which as discussed for example in~\cite{AR_method, O3_SGWB_anisotropic} is often singular and may be pseudo-inverted excluding a sub-set of eigenvalues smaller than an arbitrary threshold (i.e., conditioning). Here we follow the suggestions of~\cite{2211.10010} and perform a conditioning cut at $3\times 10^{-6}$ to invert the matrix.
    
    The map and noise estimates above can then be appropriately re-scaled to $\Ow$ units to obtain the map estimate at $f_{\rm ref}$, $\hat{\Omega}_{{\rm ref},\eta}$,
    \begin{equation}
        \hat{\Omega}_{{\rm ref},\eta} = \frac{2 \pi^2}{3 H_0^2} f_\mathrm{ref}^3 \hat{\mathcal{P}}_{\eta}.
    \end{equation}
    Maps were generated using the HEALPix pixelisation scheme \cite{astro-ph/0409513} with a resolution parameter $N_{\rm side}=16$.
    
  \subsection{Tomographic galaxy samples}\label{ssec:data.gals}
    \begin{figure}
      \centering
      \includegraphics[width=0.49\textwidth]{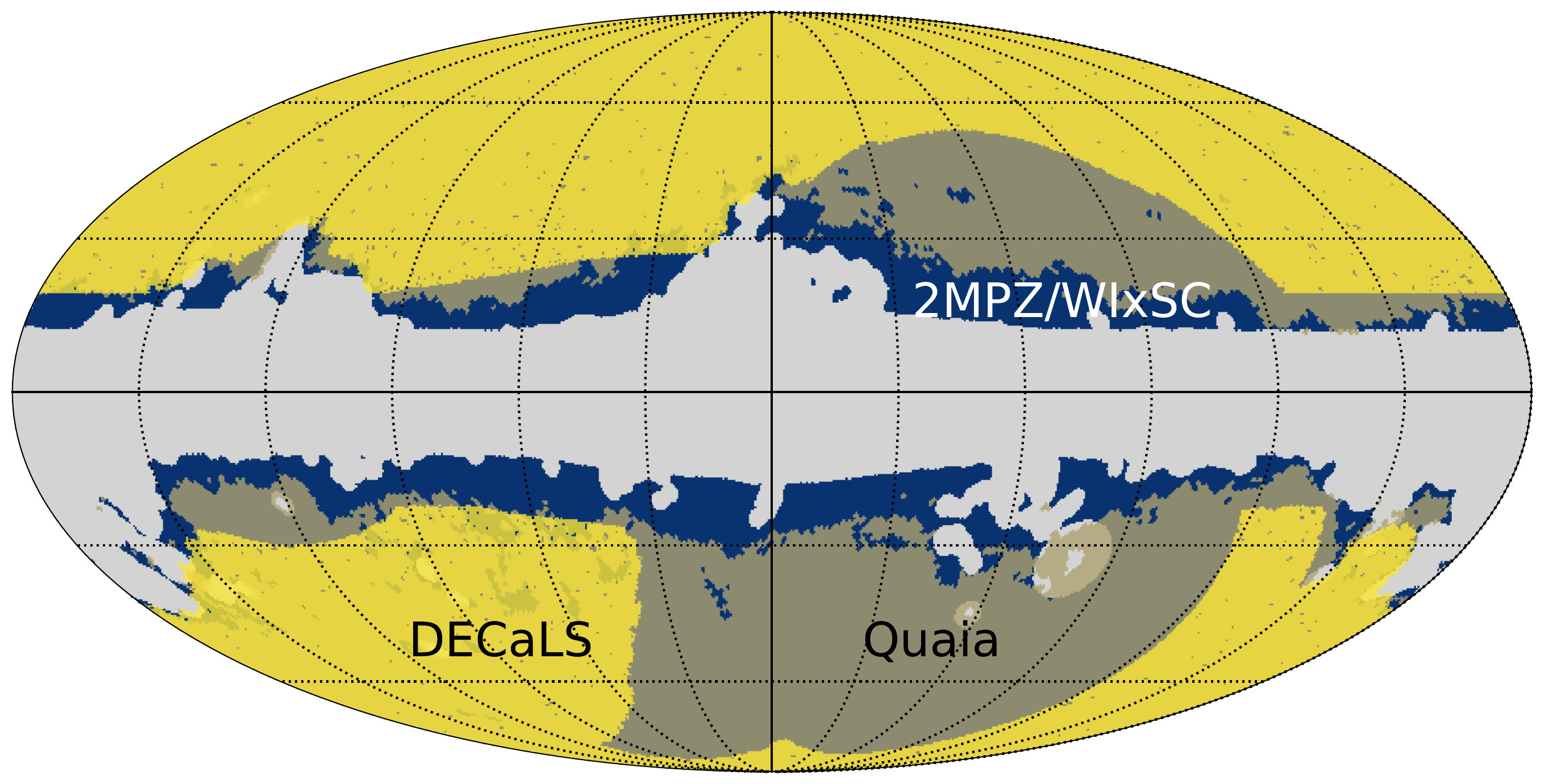}
      \caption{Sky footprints of the four galaxy samples used in this study, in Galactic coordinates. The same sky mask is assumed for \tmpz and \wisc, the selection function described in \cite{2306.17749} is used for \quaia, and the \dels mask is described in \cite{2010.00466}.}
      \label{fig:masks}
    \end{figure}
    \begin{figure}
      \centering
      \includegraphics[width=0.49\textwidth]{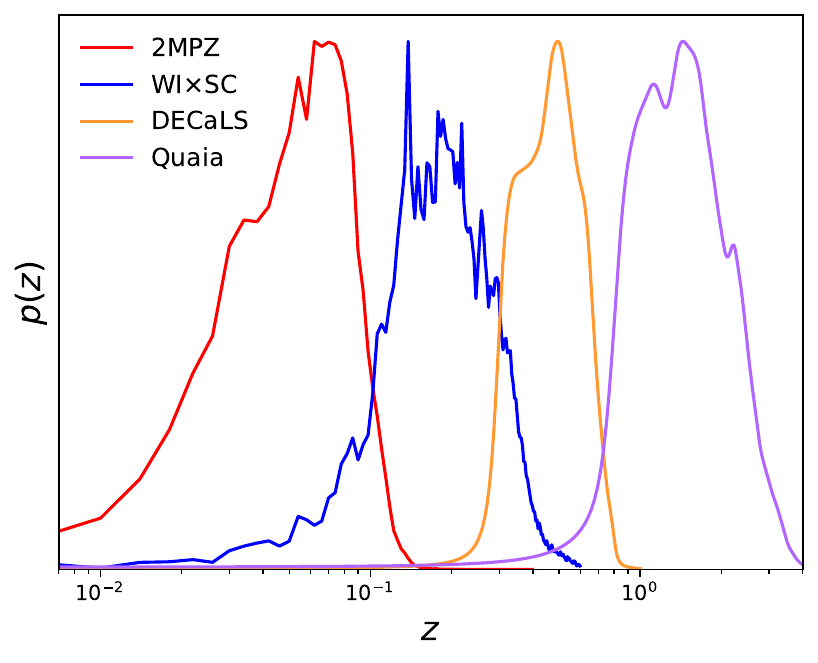}
      \caption{Redshift distributions of the four galaxy samples used in this study.}
      \label{fig:dndzs}
    \end{figure}
    We use four different galaxy samples to cross-correlate with, covering the redshift range $z\lesssim2$. These are:
    \begin{itemize}
      \item {\bf \tmpz.} The 2MASS Photometric Redshift Survey \cite{1311.5246} was constructed by matching infrared and optical photometry from the 2 Micron All-Sky Survey (2MASS, \cite{astro-ph/0004318}), SuperCOSMOS \cite{1607.01189}, and the Wide-field Infrared Survey Explorer (WISE, \cite{1008.0031}). The resulting catalog contains over 940{,}000 sources covering almost the entire celestial sphere, with relatively accurate photometric redshift accuracy ($\sigma_z\simeq0.015$), and a median redshift $z_m=0.08$. To construct our lowest-redshift sample, we select 476{,}190 2MPZ galaxies with photometric redshifts $z_p<0.1$. The sample has a mean redshift $z_{\rm \tmpz}=0.064$. We impose the sky mask used in \cite{1909.09102}, excluding the galactic plane and other sky regions heavily contaminated by dust extinction and stars. This results in a usable sky fraction $f_{\rm sky}=0.68$.
      \item {\bf \wisc.} The WISE $\times$ SuperCOSMOS Photometric Survey \cite{1607.01182}, constructed by matching the photometry from these two surveys, is $\sim3$ times deeper than \tmpz, containing about 20 million sources up to redshift $z\sim0.4$. This is achieved at the cost of less accurate photometric reshits ($\sigma_z/(1+z)\simeq0.035$), and potentially stronger contamination from stars. We correct for this as well as residual dust contamination as described in \cite{1909.09102}. We also deproject a template characterising the impact of zero-point fluctuations in the SuperCOSMOS photographic plates at the level of the galaxy auto-spectrum (again as described in \cite{1909.09102}). We select a sample of 16{,}325{,}449 galaxies in the photometric redshift range $0.1<z_p<0.4$, to which we apply the same sky mask used for \tmpz. The mean redshift of the sample is $z_{{\rm WI}\times{\rm SC}}=0.23$.
      \item {\bf \dels.} We use data from the DESI Legacy Imaging Survey \cite{1804.08657}, combining photometry from the DECam Legacy Survey \cite{1504.02900,2016AAS...22831701B}, the Mayall $z$-band Legacy Survey \cite{2016SPIE.9908E..2CD,2016AAS...22831702S}, and the Beijing-Arizona Sky Survey \cite{2004SPIE.5492..787W,1908.07099}. Specifically, we use the galaxy sample selected by \cite{2010.00466}, where each galaxy was assigned a redshift by matching them in colour space with a spectroscopic sample. We use these photometric redshifts to construct a sample in the range $0.3<z_p<0.8$, (mean redshift $z_{\rm \dels}=0.50$). The sky mask, covering 35\% of the sky, was constructed as described in \cite{2105.12108}. Sky contamination, mainly from stars, was corrected for as described in \cite{2010.00466}.
      \item {\bf \quaia.} Our highest-redshift sample is the {\sl Gaia}-unWISE quasar catalog, \quaia \cite{2306.17749}, constructed by combining photometry and spectroscopy from the {\sl Gaia} quasar sample \cite{2208.00211} and infrared data from unWISE \cite{1909.05444}. Spectro-photometric redshifts were trained on spectroscopic data from the 16th data release of the Sloan Digital Sky Survey \cite{2007.09001}, achieving a highly competitive redshift precision ($\Delta z/(1+z)<0.01$ for 63\% of the sample). We select a sample of 1{,}092{,}207 quasars with redshifts $0.8<z<5$, and a mean redshift of $z_{\rm \quaia}=1.72$. Sky contamination was corrected for using a selection function constructed as described in \cite{2306.17749}. A sky mask, covering 57\% of the sky, was constructed from this selection function as described in \cite{2306.17748}.
    \end{itemize}
    Figure \ref{fig:dndzs} shows the redshift distributions of these four samples. The sky masks associated with each of them are shown in Fig. \ref{fig:masks}.

    We construct galaxy overdensity maps for the four samples above for an angular resolution parameter $N_{\rm side}=512$. When computing the cross-correlation with the LVK GWB maps, we downgrade these maps to the same resolution $N_{\rm side}=16$. In doing so, we additionally mask out any pixel that is less than 70\% complete after downgrading (i.e. where the value of the downgraded mask is below 0.7).

\section{Data analysis methods}\label{sec:meth}
  \subsection{Power spectrum estimation}\label{ssec:meth.cls}
    Due to the different scales involved, we use different strategies to estimate the angular cross-power spectra between galaxies and the GWB maps, and the galaxy auto-spectra.

    \subsubsection{Galaxy auto-correlation}\label{sssec:meth.cls.gg}
      As described in Section \ref{ssec:theory.xcorr}, the galaxy auto-correlation of the different samples studied here is necessary in order to determine the bias of the sample. Although measures have been taken to minimise the impact of large-scale additive systematics in the overdensity maps, which would affect this auto-correlation (but not the cross-correlation), we take a conservative approach and use only the auto-spectra on angular scales $\ell>24$ (i.e. above the largest angular multipole for which we measure the GWB cross-correlation -- see the next section). The samples studied here have been found to be largely free from systematics on these scales (see \cite{1909.09102,2010.00466,2306.17748}), and this choice has the advantage that the auto- and cross-correlations can be analysed as completely independent datasets.

      Since we focus on relatively small scales, and since the noise component of the galaxy overdensity maps (namely shot noise) is largely white and homogeneous, we use the fast pseudo-$C_\ell$ method \cite{astro-ph/0105302}, as implemented in {\tt NaMaster} \cite{1809.09603}, to estimate the auto-spectrum. The power spectrum is measured in bandpowers of width $\Delta\ell=10$, in the range $\ell<3N_{\rm side}=1536$, although we impose a high-$\ell$ cut in our analysis, as described in Section \ref{ssec:meth.like}, to avoid scales where a linear bias relation becomes inaccurate. We estimate the covariance matrix of the measured power spectra using the analytical approach outlined in \cite{1906.11765}. The method requires an estimate of the unbinned true power spectrum of the data. For this we used the measured pseudo-$C_\ell$ corrected by the available sky fraction, which was found in \cite{2010.09717} to be an excellent approximation (preferable to a potentially inaccurate theoretical estimate).

    \subsubsection{Galaxy-SGWB cross-correlation}\label{sssec:meth.cls.gW}
      Current interferometric SGWB searches are characterised by a low angular resolution. In fact, it is unlikely that any current or next-generation observatory will be able to achieve a stable sensitivity beyond angular multipoles of $\ell\sim O(10)$ \cite{2005.03001}. Moreover, the resulting map-level instrumental noise is significantly non-white and inhomogeneous. Under these conditions, fast power spectrum estimators, such as the standard pseudo-$C_\ell$ approach, are highly sub-optimal, as they fail to efficiently exploit pixel-to-pixel correlations, or to weight different map modes optimally. Thus, in order to measure the galaxy-SGWB cross-correlation, we implement a quadratic minimum variance estimator, as described in \cite{astro-ph/9611174}.

      In short, in the QMVE, the data to be correlated is first inverse-variance weighted. The power spectrum of the weighted maps is then estimated, and then corrected for the impact of noise bias and mode coupling. Specifically, let ${\bf d}=({\bf m}_{\rm GW},{\bf m}_g)$ be our data, consisting of the combination of the GWB and galaxy overdensity maps, let ${\sf C}\equiv\langle {\bf d}{\bf d}^T\rangle$ be the covariance matrix of ${\bf d}$, and let ${\sf N}$ be the corresponding noise covariance matrix (i.e. the component of ${\sf C}$ due to noise in both maps). Concatenating our estimates of the power spectra into a vector ${\bf p}\equiv(C_\ell^{{\rm GW}\,{\rm GW}}, C_\ell^{{\rm GW}g}, C_\ell^{gg})$, the QMVE is:
      \begin{equation}
        \hat{\bf p}={\sf F}^{-1}(\tilde{c}-\tilde{b}),
      \end{equation}
      with $\tilde{c}$ and $\tilde{b}$ given by
      \begin{equation}
        \tilde{c}_q={\bf d}^T{\sf C}^{-1}{\sf Q}_q{\sf C}^{-1}{\bf d},\hspace{12pt}
        \tilde{b}={\rm Tr}({\sf C}^{-1}{\sf Q}_q{\sf C}^{-1}{\sf N}),
      \end{equation}
      and the Fisher matrix given by
      \begin{equation}
        F_{qq'}={\rm Tr}({\sf C}^{-1}{\sf Q}_q{\sf C}^{-1}{\sf Q}_{q'}).
      \end{equation}
      Above, the matrices ${\sf Q}_q$ are the derivatives of the covariance with respect to the power spectrum coefficients we wish to estimate (i.e. ${\sf Q}_q\equiv \partial{\sf C}/\partial p_q$). Finally, it is worth noting that, although the specific form of the estimator depends on the model used to construct the covariance matrix ${\sf C}$ (particularly the signal part of it), the estimator itself is unbiased for any choice of symmetric and positive-definite ${\sf C}$, and only its optimality depends on ${\sf C}$ being close to the true covariance of the data.

      In the case we wish to explore here, the GWB map is completely dominated by noise, and the amplitude of the cross-correlation with the galaxy overdensity should be, by comparison with the GWB noise, or the galaxy auto correlation, very small. We can therefore assume the following block-diagonal structure for the covariance matrix:
      \begin{equation}
        {\sf C}=\left(
        \begin{array}{cc}
          {\sf N}_{\rm GW} & 0 \\
          0 & {\sf C}_g
        \end{array}
        \right),
      \end{equation}
      where ${\sf N}_{\rm GW}$ is the noise covariance matrix of the GWB map, and ${\sf C}_g$ is the covariance matrix of the galaxy overdensity map (for which we provide a model below). This assumption simplifies the QMVE significantly, as it decouples the estimators for the three power spectra that make up ${\bf p}$ into three separate estimators of the form:
      \begin{equation}
        \hat{C}^{xy}_\ell=\sum_{\ell'}({\sf F}^{xy})^{-1}_{\ell\ell'}(\tilde{c}^{xy}_\ell-\tilde{b}^{xy}_\ell),
      \end{equation}
      where:
      \begin{widetext}
      \begin{align}
        &F^{{\rm GW}\,{\rm GW}}_{\ell\ell'}={\rm Tr}({\sf N}_{\rm GW}^{-1}{\sf Q}_{\ell}{\sf N}_{\rm GW}^{-1}{\sf Q}_{\ell'}),
        \hspace{12pt}\tilde{c}_\ell^{{\rm GW}\,{\rm GW}}={\bf m}_{\rm GW}^T{\sf N}_{\rm GW}^{-1}{\sf Q}_\ell{\sf N}_{\rm GW}^{-1}{\bf m}_{\rm GW},
        \hspace{12pt}\tilde{b}_\ell^{{\rm GW}\,{\rm GW}}={\rm Tr}\left({\sf N}_{\rm GW}^{-1}{\sf Q}_\ell\right), \\
        &F^{{\rm GW} g}_{\ell\ell'}={\rm Tr}({\sf N}_{\rm GW}^{-1}{\sf Q}_{(\ell}{\sf C}_{gg}^{-1}{\sf Q}_{\ell')}),
        \hspace{12pt}\tilde{c}_\ell^{{\rm GW} g}={\bf m}_{\rm GW}^T{\sf N}_{\rm GW}^{-1}{\sf Q}_\ell{\sf C}_{gg}^{-1}{\bf m}_g,
        \hspace{12pt}\tilde{b}_\ell^{{\rm GW} g}=0, \\
        &F^{gg}_{\ell\ell'}={\rm Tr}({\sf C}_{gg}^{-1}{\sf Q}_\ell{\sf C}_{gg}^{-1}{\sf Q}_{\ell'}),
        \hspace{12pt}\tilde{c}_\ell^{gg}={\bf m}_g^T{\sf C}_{gg}^{-1}{\sf Q}_\ell{\sf C}_{gg}^{-1}{\bf m}_g,
        \hspace{12pt}\tilde{b}_\ell^{gg}={\rm Tr}\left({\sf C}_{gg}^{-1}{\sf Q}_\ell{\sf C}^{-1}_{gg}{\sf N}_g\right).
      \end{align}
      \end{widetext}
      where, in the definition of ${\sf F}^{{\rm GW}g}$, the operator $(\ell\cdots\ell')$ symmetrises the two multipole indices. The ${\sf Q}_\ell$ matrices are, in turn:
      \begin{equation}
        ({\sf Q}_\ell)_{ij}=\frac{\partial C^{xy}_{ij}}{\partial C_\ell}=\frac{2\ell+1}{4\pi}P_{\ell}(\nv_i\cdot\nv_j),
      \end{equation}
      where $P_\ell(x)$ are the Legendre polynomials.

      In the equations above, the GWB noise covariance ${\sf N}_{\rm GW}$ was estimated together with the maps~\cite{2211.10010}, as described in Sec.~\ref{ssec:data.sgwb}. We construct the galaxy covariance matrix as follows:
      \begin{enumerate}
        \item We first compute a signal covariance ${\sf S}^{gg}$ as:
        \begin{equation}\label{eq:cov_gg}
          S^{gg}_{ij}=\sum_{\ell}\frac{2\ell+1}{4\pi}C_\ell^{gg}\,P_\ell(\nv_i\cdot\nv_j),
        \end{equation}
        where $C_\ell^{gg}$ is the galaxy power spectrum estimated from the high-resolution galaxy maps using a fast pseudo-$C_\ell$ estimator.
        \item Since the estimate above already accounts for the impact of shot noise (which is included in $C_\ell^{gg}$), we must only include a noise contribution for the masked pixels. We treat these as pixels with a very high noise variance, so they are suppressed in the inverse-variance-weighted map, and simply add a very large number along the elements of the diagonal corresponding to masked pixels.
      \end{enumerate}
      We verified that the covariance matrix thus constructed is positive definite and invertible, by studying its eigenvalue decomposition, and the stability of the inverse matrix under different inversion techniques\footnote{Specifically, we verified that a straightforward inverse of the form $({\sf S}+{\sf N})^{-1}$, where ${\sf N}$ is the noise covariance described, achieves the same result as the potentially more stable expression ${\sf N}^{-1}({\sf S}^{-1}+{\sf N}^{-1})^{-1}{\sf S}^{-1}$, which only depends on the individual inverses of ${\sf S}$ and ${\sf N}$.}.
      \begin{figure}
        \centering
        \includegraphics[width=0.49\textwidth]{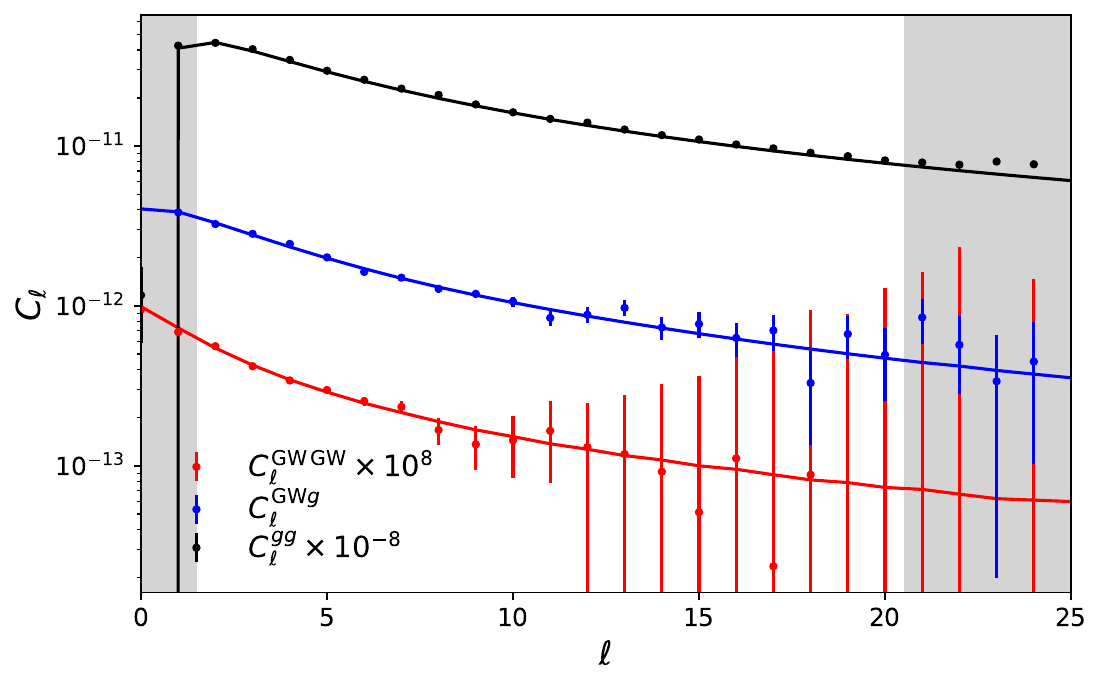}
        \caption{Angular power spectra estimated from 1000 Gaussian simulations. Results are shown in black, blue, and red for the galaxy auto-correlation, the galaxy-GWB cross-correlation, and the GWB auto-correlation, respectively. The redshift distribution and galaxy bias of \tmpz were used for this study, and correlated GWB signal with $\bOw=10^{-8}$ was injected in the data. The points with error bars show the mean and its error (i.e. scaled by $1/\sqrt{N_{\rm sims}}$), while the solid lines show the input spectra, demonstrating that our estimator is unbiased on the range of scales used for the analysis.}
        \label{fig:cl_val}
      \end{figure}

      We apply this estimator to the low-resolution ($N_{\rm side}=16$) GWB and galaxy overdensity maps described in Section \ref{sec:data}, and reconstruct the power spectrum at all integer multipoles in the range $\ell\in[0,24]$. It is worth noting that, in this analysis, we only use the QMV estimator to measure the GWB-galaxy cross-correlation, although we have outlined the general formalism above for completeness. We validate our implementation of the method in two different ways. First, we used it to estimate the GWB auto-correlation, and verified that the result was compatible with the measurements presented in \cite{2211.10010}.
      \begin{figure*}
        \centering
        \includegraphics[width=0.9\textwidth]{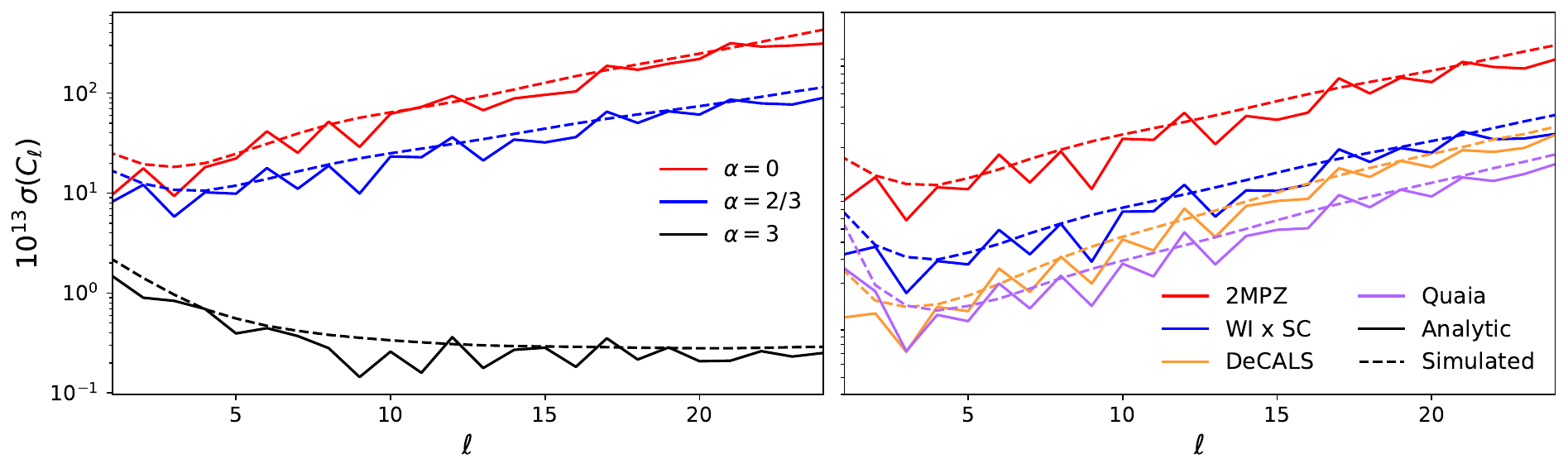}
        \caption{Statistical uncertainties in the estimated cross-spectrum $C_\ell^{{\rm GW}g}$. Results are shown as solid and dashed lines for the simulated and analytical covariances, respectively. The left panel shows the result for the \tmpz galaxy sample and the three different values of $\alpha$ explored here. The right panel shows the result for the $\alpha=2/3$ map and for the 4 different galaxy samples. As expected, the $\alpha=3$ map exhibits better noise properties on smaller scales, owing to the larger weight placed on the higher frequencies. We find that, although the analytical error estimates are a reasonable representation of the underlying uncertainties, they are mildly over-estimated when compared with the simulations. This motivates the use of the hybrid covariance matrix described in Section \ref{sssec:meth.cls.covx}.}
        \label{fig:cov_std}
      \end{figure*}
      
      To more thoroughly validate our implementation, we tested it against a suite of 1000 simulations with known input power spectra. Each simulation consisted of two signal-only maps, for the GWB and the galaxy overdensity, generated as correlated Gaussian random fields with theory power spectra $(C^{{\rm GW}\,{\rm GW}}_\ell,C^{{\rm GW}g}_\ell,C_\ell^{gg})$ calculated as described in Section \ref{sec:theory}, assuming the redshift distribution and galaxy bias of the \tmpz sample, and a bias-weighted GW production rate $\bOw=10^{-8}\,{\rm Gyr}^{-1}$. To this we added a Gaussian GWB noise realisation drawn from the noise covariance ${\sf N}_{\rm GW}$ for the GWB map with $\alpha=2/3$. We then applied the same angular mask, and estimated all power spectra using the QMVE algorithm described above. Fig. \ref{fig:cl_val} shows the mean power spectra of the 1000 simulations (points with error bars), compared with the input power spectra. Note that the error bars denote the error in the mean (i.e. the scatter between the 1000 simulations divided by $\sqrt{1000}$). Our QMVE implementation is able to obtain an unbiased estimate of all power spectra.

    \subsubsection{Cross-correlation covariance}\label{sssec:meth.cls.covx}
      \begin{figure}
        \centering
        \includegraphics[width=0.49\textwidth]{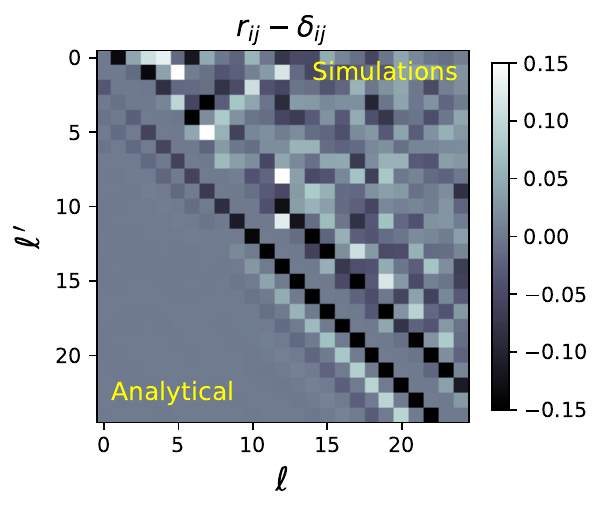}
        \caption{Correlation matrix of the galaxy-GWB cross-spectrum for the \tmpz sample and $\alpha=2/3$. The lower triangular part shows the structure of the analytical covariance, while the noisier simulated result is shown in the upper-triangular sector. The diagonal elements of the matrix (which are 1 by definition) have been set to zero to better visualise the correlation structure. }
        \label{fig:cov_ana_sim}
      \end{figure}
      \begin{figure*}
        \centering
        \includegraphics[width=0.9\textwidth]{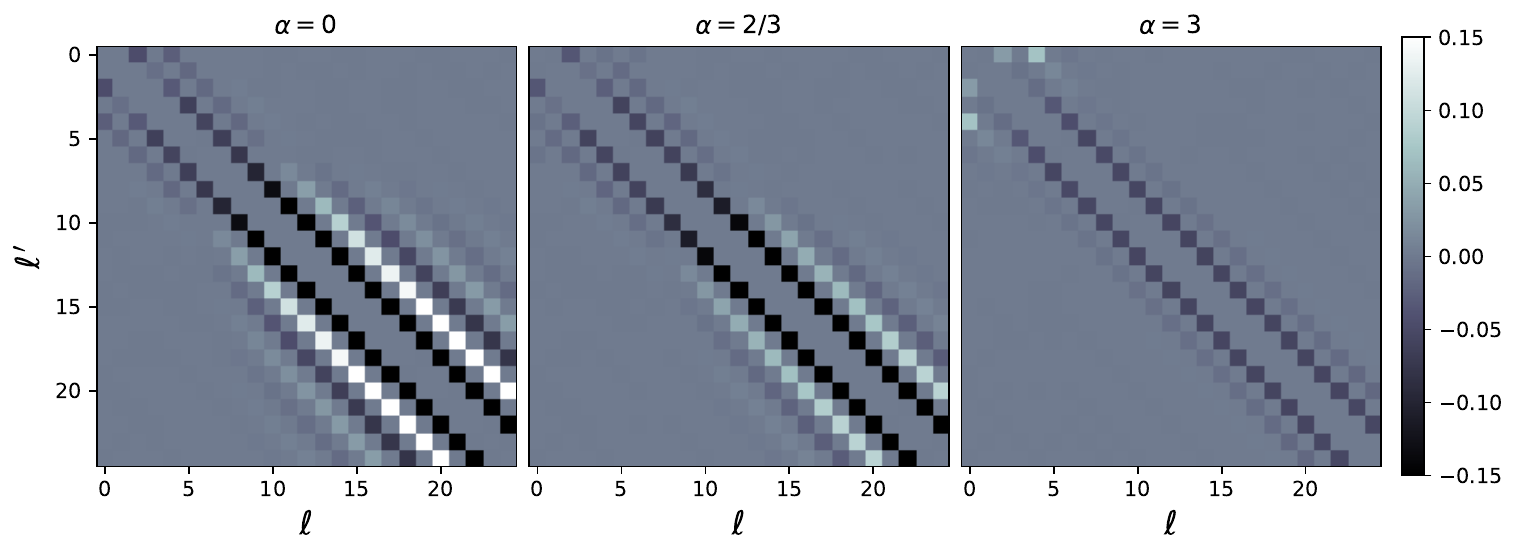}
        \caption{Correlation matrix of the galaxy-GWB cross-spectrum for \tmpz and the three different values of $\alpha$ explored here. The diagonal elements of the matrix (which are 1 by definition) have been set to zero to better visualise the correlation structure. The noise properties of the different GWB maps, connected with the weight assigned to differeng GW frequencies, has an impact on the specific correlation structure of the measurements.}
        \label{fig:cov_ralpha}
      \end{figure*}
      In addition to the cross-power spectrum itself, we must also estimate its uncertainties, namely its covariance matrix. Assuming all fields to be Gaussian, and assuming that ${\sf C}$ is a sufficiently accurate estimate of the map-level covariance, the power spectrum covariance would be simply given by the inverse of the fisher matrix ${\sf F}$. Any inaccuracy in the estimated galaxy overdensity map covariance (e.g. due to the band limit used in Eq. \ref{eq:cov_gg}), or any over- or under-estimate of the GWB noise covariance (which could be due to the regularising eigenvalue threshold described in Section \ref{ssec:data.sgwb}), would lead to a misestimation of the power spectrum uncertainties. To avoid this, we construct an alternative estimate of the covariance matrix from simulations. In detail, for each galaxy sample we construct a suite of 1000 Gaussian realisations of the overdensity map following the measured auto-power spectrum. We then compute the cross-spectra of all galaxy realisations with the real GWB map, and estimate the covariance matrix as
      \begin{equation}\nonumber
        {\rm Cov}(C^{{\rm GW}g}_\ell,C^{{\rm GW}g}_{\ell'})=\frac{\sum_{i=1}^{N_{\rm sim}}(C^{{\rm GW}g,i}_\ell-\bar{C}^{{\rm GW}g}_\ell)(C^{{\rm GW}g,i}_{\ell'}-\bar{C}^{{\rm GW}g}_{\ell'})}{N_{\rm sim}-1},
      \end{equation}
      where $C^{{\rm GW}g,i}_\ell$ is the power spectrum of the $i$-th simulation, and $\bar{C}^{{\rm GW}g}_\ell$ is the average over simulations. Through this method, we are intentionally neglecting any contribution to the power spectrum covariance arising from a genuine correlation between the GWB map and the galaxy overdensity field, since our simulations are not correlated with the data. However, since any such correlation, as we will see a posteriori, is negligible compared to the variance in the GWB and galaxy maps, this is a reasonable approximation. In turn, this method allows us to incorporate into the estimated covariance the true noise properties of the GWB map, without relying on the noise covariance matrix ${\sf N}_{\rm GW}$.

      Fig. \ref{fig:cov_std} shows the power spectrum uncertainties estimated using this method (solid lines) together with the analytical estimate constructed from the Fisher matrix of the QMVE (dashed lines) for different combinations of galaxy sample and GWB spectral index $\alpha$. We find that, although both estimates agree with each other to a large extent, the simulation-based method seems to be consistently smaller than the analytical uncertainties, particularly at low $\ell$, indicating a potential over-estimation of the map-level GWB noise covariance. Fig. \ref{fig:cov_ana_sim} shows the correlation matrix $r_{\rm ij}\equiv {\rm Cov}_{ij}/\sqrt{{\rm Cov}_{ii}{\rm Cov}_{jj}}$, which quantifies the correlation between the uncertainties in different multipoles, for both methods (in the upper and lower-triangular parts of the matrix). We see that, although both estimates reproduce a similar correlation structure, the simulation-based matrix is significantly noisier, in spite of the large number of simulations used to estimate it. For this reason, we build the final covariance matrices combining the simulation-based and analytical approaches as:
      \begin{equation}
        {\rm Cov}_ij={\rm Cov}_{ij}^{\rm A}\sqrt{\frac{{\rm Cov}_{ii}^{\rm S}{\rm Cov}_{jj}^{\rm S}}{{\rm Cov}_{ii}^{\rm A}{\rm Cov}_{jj}^{\rm A}}},
      \end{equation}
      preserving the correlation structure of the analytical covariance but employing the diagonal errors estimated via simulations. Fig. \ref{fig:cov_ralpha}, which shows the correlation matrix for the three different values of $\alpha$ of the GWB maps, demonstrates that this correlation structure is largely governed by the relative contribution of different frequencies to the final GWB map. This is as expected, since the angular scales to which a given GW interferometer with fixed baselines is sensitive depend strongly on the GW frequency~\cite{AR_O1O2}.

  \subsection{Likelihoods and parameter inference}\label{ssec:meth.like}
    \begin{figure*}
      \centering
      \includegraphics[width=0.7\textwidth]{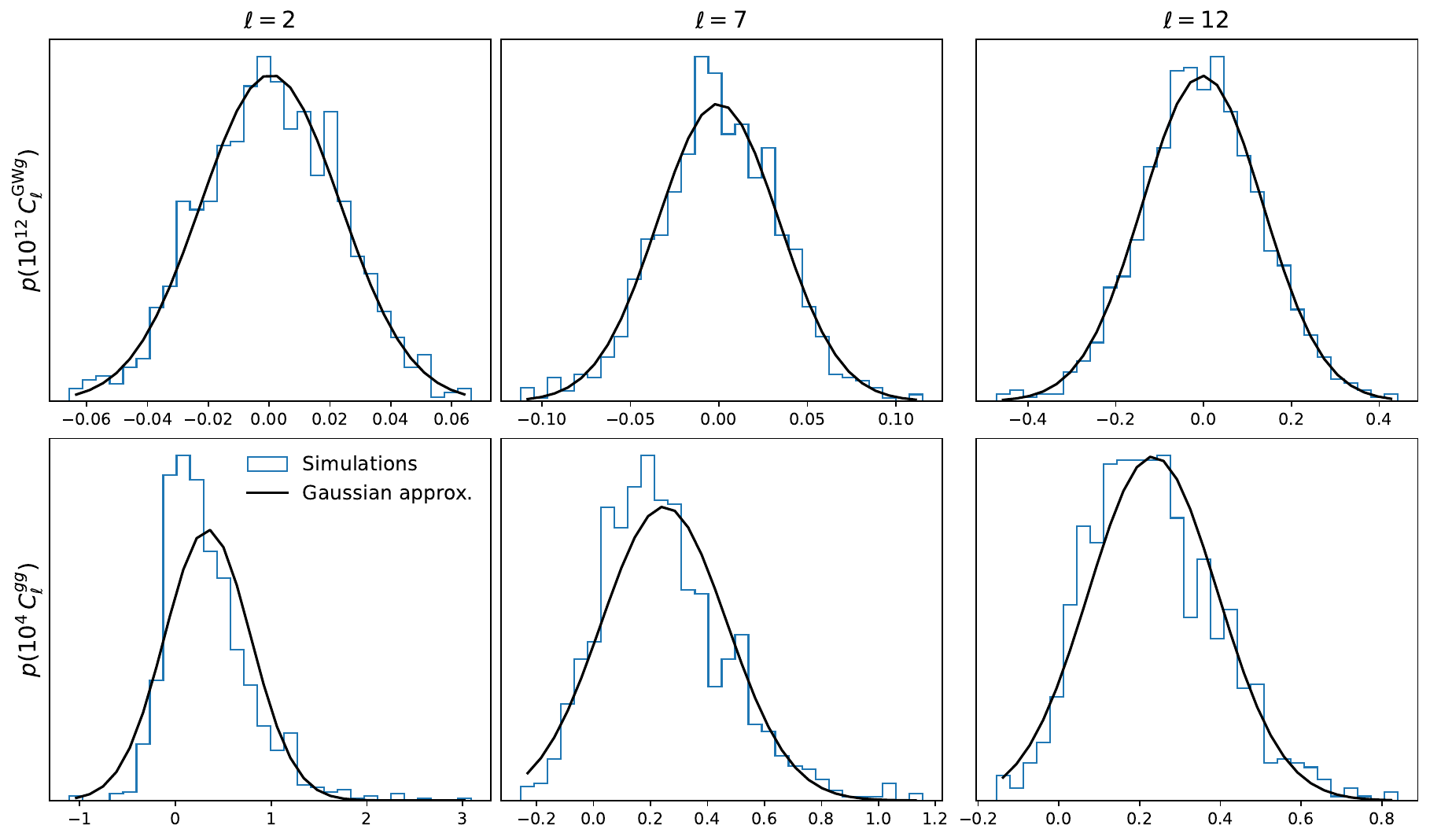}
      \caption{Distribution of the measured power spectrum multipoles estimated from simulations (blue histogram), together with its Gaussian approximation (black solid line). Results are shown for the galaxy-GWB cross-correlation (top row) and for the galaxy auto-correlation (bottom), for the \tmpz sample and $\alpha=2/3$. Although the Gaussian approximation fails at low $\ell$ in the case of the auto-correlation, it reproduces the true distribution of the cross-correlation multipoles at high accuracy.}\label{fig:cllike}
    \end{figure*}
    In order to extract parameter constraints from the measured angular power spectra we must first construct the likelihood $p({\bf d}|\vec{\theta})$, describing the distribution of possible values for the measured data vector ${\bf d}$ for a given set of model parameters $\vec{\theta}$. In our case, $\vec{\theta}=(b_g,\bOw)$, and the data vector ${\bf d}\equiv\{\vec{C}^{gg}_\ell,\vec{C}^{{\rm GW}g}_\ell\}$ comprises measurements of the galaxy auto-power spectrum in bandpowers of $\Delta\ell$ at $\ell>24$, and measurements of the galaxy-GWB cross-correlation at all integer $\ell$s in the range $\ell\in[\ell_{\rm min},20]$ (see Section \ref{ssec:res.cls}). Since both components of the data vector cover disjoint multipole ranges, we can treat them as independent variables to a good approximation, such that
    \begin{equation}
      p({\bf d}|\vec{\theta})=p(\vec{C}^{gg}_\ell|b_g^2)\,p(\vec{C}^{{\rm GW}g}_\ell|b_g\bOw),
    \end{equation}
    where we have explicitly noted that $C^{gg}_\ell$ and $C^{{\rm GW}g}_\ell$ only depend on the parameter combinations $b_g^2$ and $b_g\bOw$, respectively.

    At the relatively high $\ell$s for which we use the galaxy autocorrelation, the central limit theorem ensures that $p(C_\ell^{gg}|b_g^2)$ is well approximated by a multivariate normal distribution with a fixed covariance matrix (estimated as described in Section \ref{sssec:meth.cls.gg}).
    
    The same argument cannot, in principle, be used to derive the likelihood of the low-$\ell$ cross-correlation $C_\ell^{{\rm GW}g}$. Indeed, assuming a given field $m(\nv)$ to be Gaussianly distributed, since its auto-power spectrum involves a quadratic combination of $m$, the power spectrum likelihood is better described by a Wishart distribution \cite{0801.0554}, which can depart significantly from a Gaussian at low $\ell$. This can be seen explicitly in the bottom row of Fig. \ref{fig:cllike}, which shows the distribution of galaxy auto-power spectra in the Gaussian simulations used to estimate the cross-correlation covariance as described in Section \ref{sssec:meth.cls.covx}. Results are shown for the \tmpz sample and the $\alpha=2/3$ GWB map. As shown in the Figure, the distribution of auto-power spectrum multipoles (blue histogram) departs visibly from the Gaussian approximation (solid black line) at low $\ell$, and approaches it at $\ell\gtrsim12$ (which validates the Gaussian assumption made for the high-$\ell$ auto-spectrum described above). The top row of the same figure shows the distribution of galaxy-GWB cross-spectrum multipoles. Interestingly, in this case we find that the distribution of spectra follows the Gaussian approximation at very high accuracy at all $\ell$s (down to $\ell=2$). Although this result may be a priori surprising, we can understand it as follows: if the statistical uncertainty in a cross-correlation is dominated by the instrumental noise of one of the fields being correlated (as is the case of the GWB maps used here), the cross-correlation estimator can be considered to depend linearly on that field, while the second field is effectively fixed (since its variance is much smaller). Since Gaussianity is preserved under linear transformations, it is therefore natural that the cross-spectra should follow a distribution that is close to Gaussian. In summary, we find that the cross-spectrum likelihood $p(C_\ell^{{\rm GW}g}|b_g\bOw)$ is well described as a multivariate Gaussian, with the covariance described in Section \ref{sssec:meth.cls.covx}.
      \begin{figure*}
        \centering
        \includegraphics[width=0.49\textwidth]{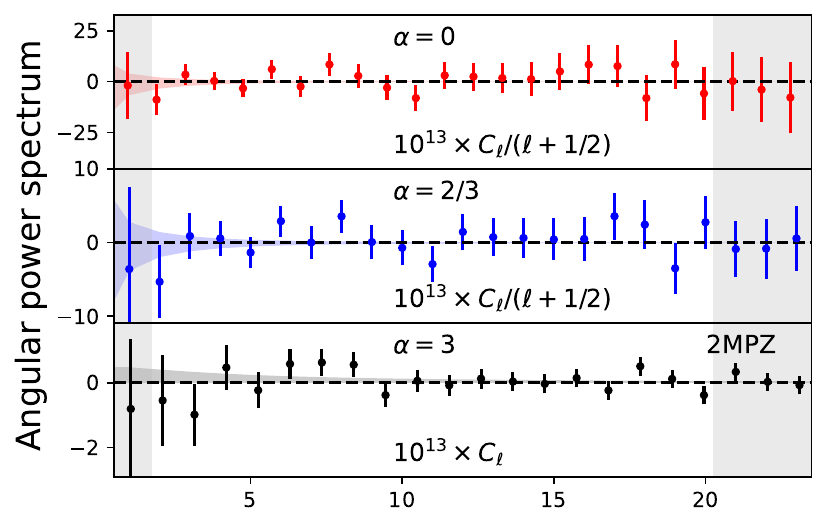}
        \includegraphics[width=0.47\textwidth]{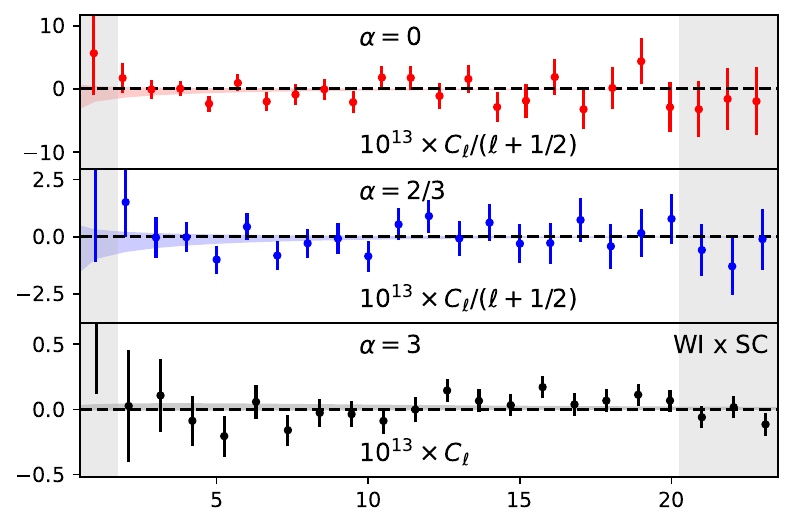}
        \includegraphics[width=0.48\textwidth]{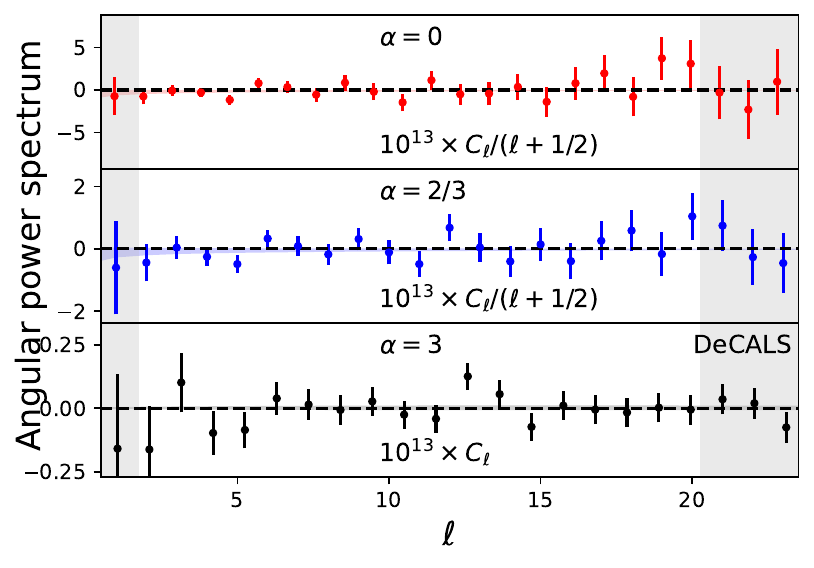}
        \includegraphics[width=0.47\textwidth]{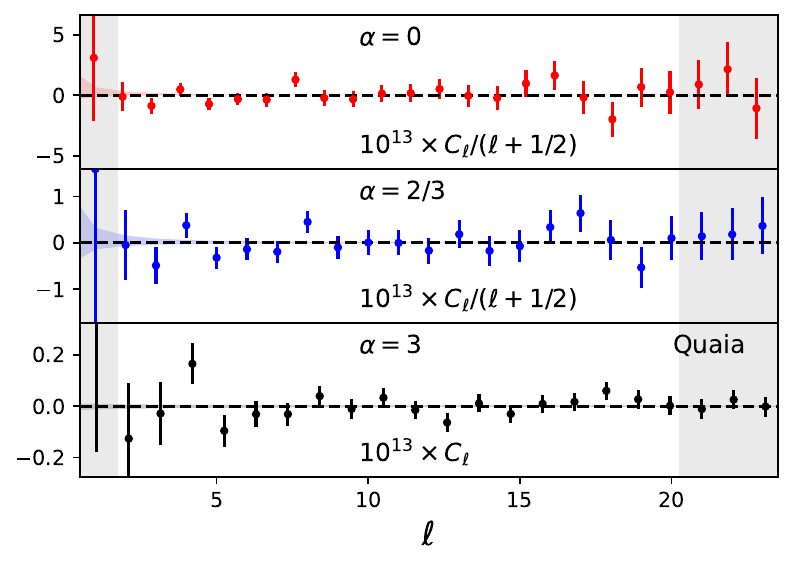}
        \caption{Angular cross-spectra between the four galaxy samples (\tmpz, \wisc, \dels, \quaia, from left to right and top to bottom), and the three different values of $\alpha$ (0, 2/3, and 3 from top to bottom in each panel). In each panel, the gray vertical bands show the angular multipoles excluded from the fiducial analysis. The coloured shaded bands show the $1\sigma$ intervals allowed by the constraints on $\bOw$ according to the theoretical predictions of Section \ref{ssec:theory.xcorr}.}
        \label{fig:cls}
      \end{figure*}

    {

    The auto- and cross-spectrum likelihoods thus follow the same structure:
    \begin{equation}\label{eq:linpar}
      -2\log p({\bf d}_\alpha|b_\alpha)=({\bf d}_\alpha-b_\alpha{\bf t}_\alpha)^T{\sf C}_\alpha^{-1}({\bf d}_\alpha-b_\alpha{\bf t}_\alpha)+{\rm const.},
    \end{equation}
    where $\alpha\in\{gg,{\rm GW}g\}$, the data vector ${\bf d}_\alpha$ is the measurements of either $C_\ell^{gg}$ or $C_\ell^{{\rm GW}g}$, ${\sf C}_\alpha$ is the relevant covariance matrix, $b_\alpha\in\{b_g^2,b_g\bOw\}$, and ${\bf t}_\alpha\in\{T_\ell^{gg},T_\ell^{{\rm GW}g}\}$ are the fixed templates of Eqs. \ref{eq:tgg} and \ref{eq:tgw}. Assuming a flat prior on the linear parameters $b_\alpha$, their posterior distribution is therefore Gaussian, with a mean and variance given by:
    \begin{equation}
      \bar{b}_\alpha=\frac{{\bf t}_\alpha^T{\sf C}^{-1}_\alpha{\bf d}_\alpha}{{\bf t}_\alpha^T{\sf C}^{-1}_\alpha{\bf t}_\alpha},\hspace{12pt}
      \sigma(\bar{b}_\alpha)=\left({\bf t}_\alpha^T{\sf C}^{-1}_\alpha{\bf t}_\alpha\right)^{-1/2}.
    \end{equation}
    Once $b_g^2$ and $b_g\bOw$, and their variances, have been estimated, the marginalised constraints on $\bOw$ (the parameter we are interested in), could be obtained from a fast Monte-Carlo chain of the corresponding two-dimensional Gaussian posterior distribution over $(b_g,\bOw)$. In practice, however, the uncertainties on $b_g^2$ from the galaxy auto-correlation are much smaller than the uncertainties on $b_g\bOw$ (which thus dominate the final $\bOw$ error). We thus find that simple linearised error propagation captures the mean and variance of $\bOw$ to better than 0.2\%. In short, let $x\equiv b_g^2$ and $y\equiv b_g\bOw$. Then, the posterior mean and error on $\bOw$ are, to a very good approximation:
    \begin{align}\nonumber
      \overline{\bOw}=\frac{\bar{y}}{\sqrt{\bar{x}}},\hspace{12pt}\sigma(\bOw)=\left[\frac{\sigma^2(y)}{\bar{x}}+\frac{\sigma^2(x)}{4}\frac{\bar{y}^2}{\bar{x}^3}\right]^{1/2}.
    \end{align}

\section{Results}\label{sec:res}
  \subsection{Power spectrum measurements}\label{ssec:res.cls}
    \renewcommand{\arraystretch}{1.5}
    \setlength{\tabcolsep}{5pt}
    \begin{table*}
      \centering
      \begin{tabular}[t]{|c|c|c|c|c|c|c|c|}
        \hline
        \multirow{2}{*}{Sample} & \multirow{2}{*}{$\bar{z}$} &
        \multicolumn{3}{c|}{$10^8\times\bOw\,\,[{\rm Gyr}^{-1}]$} &
        \multicolumn{3}{c|}{${\rm PTE}(C_\ell^{{\rm GW}g})$} \\ \cline{3-8}
        & & $\alpha=0$ & $\alpha=2/3$ & $\alpha=3$ & $\alpha=0$ & $\alpha=2/3$ & $\alpha=3$ \\
        \hline
        \hline
        \tmpz & 0.06  & $-0.057\pm 0.288$  & $0.004\pm 0.161$  & $0.007\pm 0.007$  & 0.875  & 0.868  & 0.536 \\
        \wisc & 0.23  & $-0.679\pm 1.142$  & $-0.277\pm 0.587$  & $0.011\pm 0.015$  & 0.625  & 0.844  & 0.604 \\
        \dels & 0.50  & $-8.809\pm 9.231$  & $-3.233\pm 4.567$  & $0.023\pm 0.086$  & 0.632  & 0.651  & 0.677 \\
        \quaia & 1.73  & $-24.172\pm 31.667$  & $-8.591\pm 16.880$  & $-0.028\pm 0.491$  & 0.612  & 0.594  & 0.488 \\
        \hline
      \end{tabular}\label{tab:zbins}
      \caption{Constraints (posterior mean and 68\% Gaussian errors) on $\bOw$ for the four tomographic galaxy samples and the three different values of $\alpha$. The last three columns show the probability-to-exceed (PTE) of the $\chi^2$ with respect to the null hypothesis for each power spectrum. We find no significant detection of the cross-correlation.}
    \end{table*}
    }

    We measure the cross-correlation between the GWB maps constructed with three different frequency scalings ($\alpha\in\{0,2/3,3\}$), and maps of the galaxy overdensity in our 4 tomographic samples (\tmpz, \wisc, \dels, and \quaia, in order of ascending mean redshift), for a total of 12 power spectra. These are constructed using the optimal quadratic estimator described in Section \ref{ssec:meth.cls}, and measured at all integer multipoles in the range $\ell\in[0,24]$. As shown in Fig. \ref{fig:cl_val}, the power spectra can be recovered accurately up to $\ell=20$, with a small bias due to edge effects at higher multipoles. Since the noise sensitivity drops sharply on small scales, we simply discard these multipoles. Since the monopole of the galaxy overdensity maps is zero by definition, we discard the $\ell=0$ mode. Additionally, in order to avoid potential contamination from a non-cosmological dipole component in the galaxy density map, which may not be well described by our covariance matrix, we also discard the $\ell=1$ mode in our fiducial analysis. These cuts were selected before carrying out the analysis, and not imposed a posteriori. In fact, we do not detect any significant ($>2\sigma$) dipole or monopole in our measurements. Our fiducial constraints, shown below (see Table \ref{tab:zbins}) are not significantly affected by these choices, and only change by a fraction of the statistical uncertainties when including $\ell\in\{0,1\}$.
      \begin{figure*}
        \centering
        \includegraphics[width=0.7\textwidth]{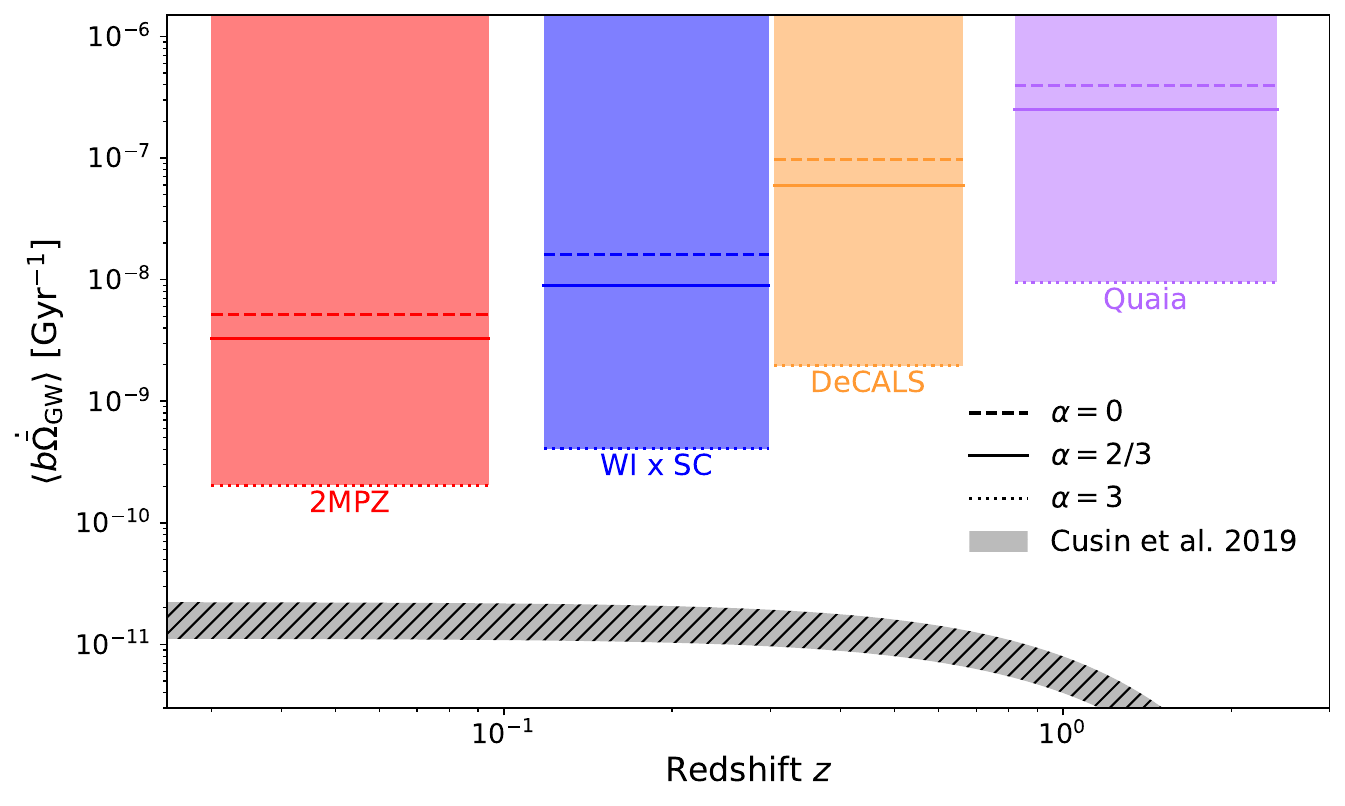}
        \caption{Tomographic 95\% upper bounds on the bias-weighted production rate of gravitational waves as a function of redshift for different values of the frequency scaling parameter $\alpha$. The expected astrophysical signal, according to the Reference model of \cite{1904.07797} is shown as a gray hatched band.}
        \label{fig:bOw}
      \end{figure*}
      \begin{figure}
        \centering
        \includegraphics[width=0.49\textwidth]{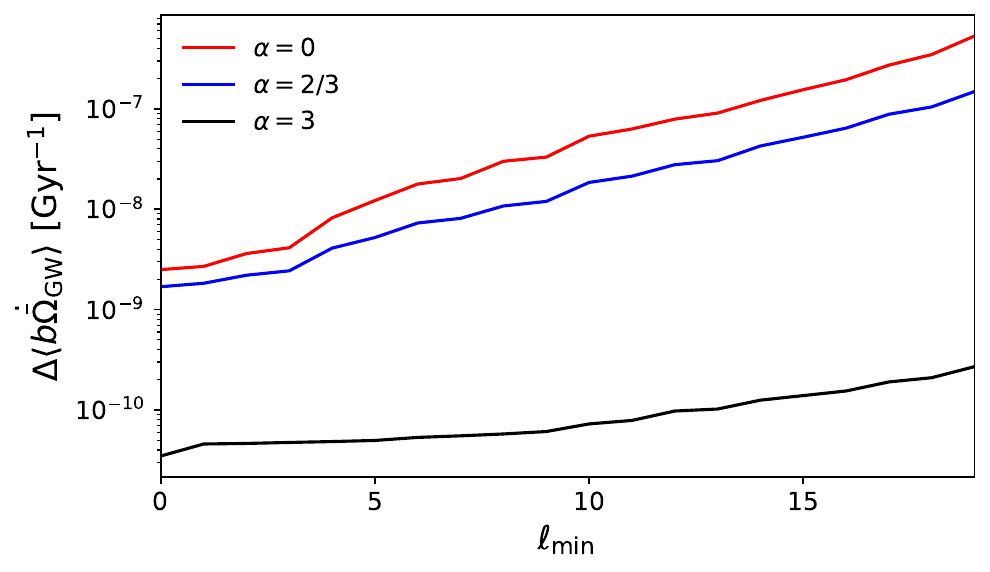}
        \caption{Dependence of the statistical uncertainties of our $\bOw$ measurements on the minimum multipole included in the analysis. The cases $\alpha=0$ and $\alpha=2/3$ show a steeper increase with $\ell_{\rm min}$ than the $\alpha=3$ map, which recovers more information from smaller scales.}
        \label{fig:lmin}
      \end{figure}

    The measured power spectra are shown in Fig. \ref{fig:cls} for \tmpz, \wisc, \dels, and \quaia, from left to right and top to bottom. Results are shown in red, blue, and black, for $\alpha=0$, 2/3, and 3, respectively. In the case of $\alpha=0$ and 2/3, we show the power specra scaled by $(\ell+1/2)^{-1}$, in order to display all points and error bars on the same linear scale (the map-level noise for these frequency scalings rises significantly faster on small scales than in the $\alpha=3$ case). In all cases, we find no significant detection of the cross-correlation visually. To quantify this, we calculate the $\chi^2$ of these measurements with respect to the null hypothesis (i.e. $\chi_0^2\equiv {\bf d}_{{\rm GW}g}^T{\sf C}_{{\rm GW}g}^{-1}{\bf d}_{{\rm GW}g}$, in the notation of Section \ref{ssec:meth.like}), and the associated probability-to-exceed (PTE). These PTEs are listed in the last three columns of Table \ref{tab:zbins}. As expected, we find no significant evidence of a cross-correlation, with all PTEs values around or above 50\%.

  \subsection{Tomographic constraints on the astrophysical GW production rate}\label{ssec:res.tomo}
    Although our measurements show no evidence of a cross-correlation between the LVK maps and the large-scale structure, we can use them to place an upper-bound constraint on the bias-weighted production rate $\bOw$ at the mean redshift of each of our samples. The resulting constraints on $\bOw$ are shown in Table \ref{tab:zbins}. Although an astrophysical background of merging binaries would most likely be described by a frequency spectrum with $\alpha\simeq2/3$, we obtain results for $\alpha=0$ and $\alpha=3$ too. Fig. \ref{fig:bOw} shows the 2$\sigma$ upper bounds found in each redshift bin (calculated as $\overline{\bOw}+2\sigma(\bOw)$, where $\overline{\bOw}$ is the posterior mean and $\sigma(\bOw)$ is its standard deviation).

    In all cases, we find tighter bounds for the $\alpha=3$ map. This is expected, since the larger weight of high frequencies significantly reduces the noise level on the smaller scales probed here, which we can then exploit to measure $\bOw$. As mentioned above, the most likely astrophysical background that would correlate with the large-scale structure (merging binary systems) would however be closer to the $\alpha=2/3$ scaling, for which our constraints are a factor $\sim40$ weaker. To illustrate this point, Fig. \ref{fig:lmin} shows the dependence of the $\bOw$ error on the minimum multipole $\ell_{\rm min}$ used in the analysis. We observe a steeper growth after $\ell_{\rm min}\sim 4$ for $\alpha=0$ and $\alpha=2/3$ than in the case of $\alpha=3$, which obtains smaller measurement uncertainties on smaller scales.

    The lack of a detection in our measurements is not particularly surprising. The gray hatched band shown in Fig. \ref{fig:bOw} corresponds to the value of $\bOw(z)$ predicted by the ``Reference'' astrophysical model of \cite{1904.07797}. This prediction was calculated using the fitting function presented in \cite{2304.07621}. Note that, strictly speaking, the model only predicts $\langle\dot{\Omega}_{\rm GW}\rangle$, and therefore the gray band shows the range of possible values allowing for the mean luminosity-weighted bias of GW emitters to vary in the range $b_{\rm GW}\in[1, 2]$). The predicted signal is more than two orders of magnitude lower than our current constraints.
  
  \subsection{Future forecasts}\label{ssec:res.forecasts}
    While the current sensitivity of stochastic GWB maps is not high enough to probe the range of signals expected from current astrophysical models, it is worth looking to the future to quantify the potential of next-generation experiments to test these models. Key in this calculation is predicting the map-level noise properties of any given network of detectors given their detector sensitivities. For this, we use the public package {\tt schNell}\footnote{\url{https://github.com/damonge/schNell}} \cite{2005.03001}, which computes the noise power spectrum $N_\ell$ of GWB maps built from a given configuration of gravitational-wave detectors. In particular, we consider two networks:
    \begin{itemize}
      \item A network consisting of the two LIGO detectors, as well as Virgo and KAGRA, in combination with a next-generation experiment with a sensitivity similar to the Einstein Telescope (ET) \cite{0810.0604}. We use the publicly available noise PSD curves for these experiments, and consider a triangular configuration for the ET-like site, as described in \cite{2005.03001}.
      \item A network consisting of two next-generation experiments, modelled after ET and the Cosmic Explorer (CE) \cite{1907.04833}. We use the publicly available PSD curve for CE, assuming its baseline configuration.
    \end{itemize}
    In the first case (LVK+ET), we consider only cross-correlations between pairs of different detectors. For ET, with a triangular configuration, this implies considering only correlations between any given triangle vertex and the other two, as well as any other detector in the network. We discard detector auto-correlations due to the difficulty in subtracting the contribution from noise correlations at the map level in a reliable manner. In the second case (ET+CE), we take all auto- and cross-correlations into account. The auto-correlations can be used to significantly boost the sensitivity to the $\ell=2$ and $\ell=4$ modes, as shown in \cite{2005.03001}. Therefore, this represents an optimistic scenario, in which mapping techniques have been designed to reliably recover the signal from detector auto-correlations.
      \begin{figure*}
        \centering
        \includegraphics[width=0.7\textwidth]{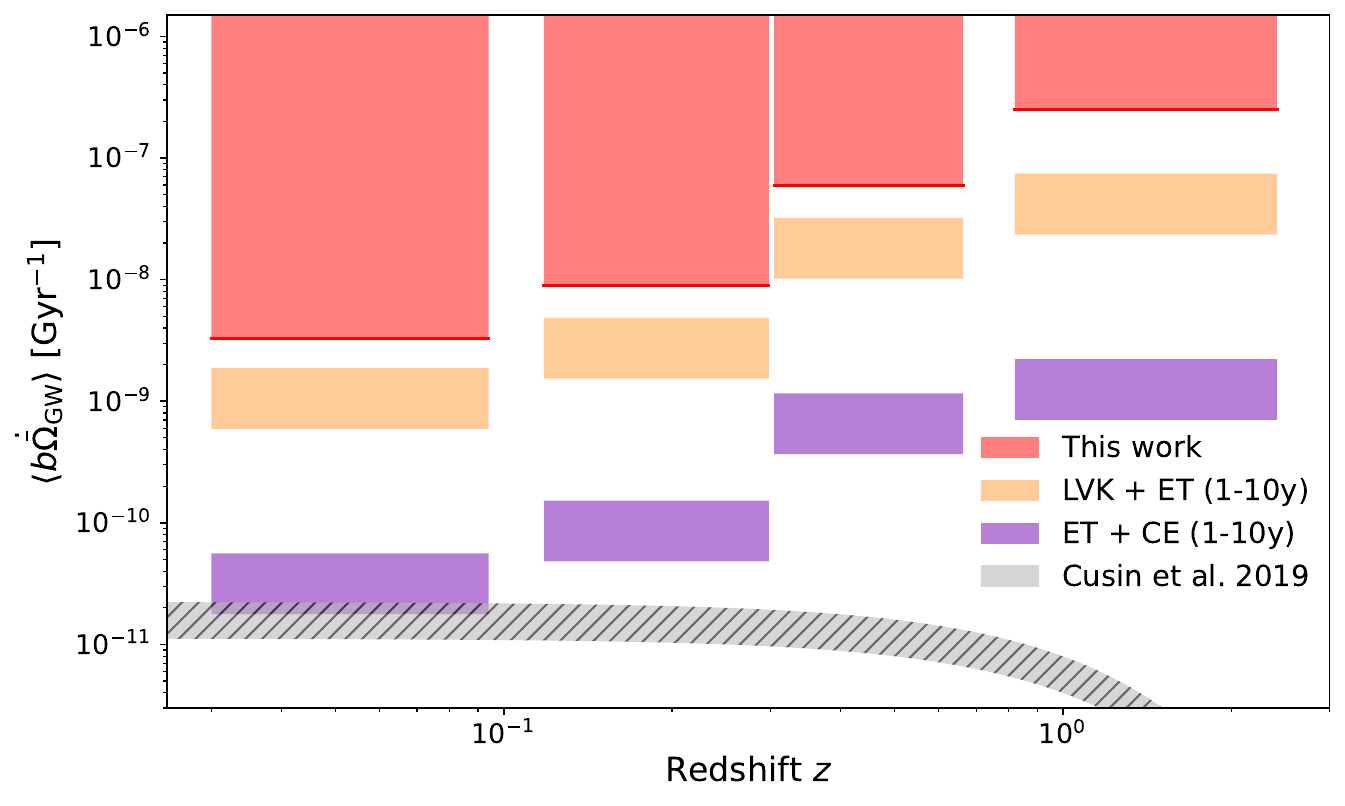}
        \caption{Current 2$\sigma$ upper bounds (red) on the bias-weighted GW production rate $\bOw$, and forecast constraints for future experiments, including a combination of LIGO-Virgo-KAGRA with an Einstein Telescope-like experiment (yellow), and a combination of two next-generation observatories, such as the Einstein Telescope and the Cosmic Explorer (purple). The expected astrophysical signal, according to the Reference model of \cite{1904.07797} is shown as a gray hatched band.}
        \label{fig:bOw_4cast}
      \end{figure*}

    After estimating the $N_\ell$, we calculate the associated uncertainty in $\bOw$ using Eq. \ref{eq:linpar}, with a theory template ${\bf t}_{{\rm GW}g}$ computed as described in Section \ref{ssec:theory.xcorr}. We construct the covariance matrix assuming purely Gaussian statistics and full-sky observations as \cite{astro-ph/9606066}
    \begin{align}\nonumber
      {\rm Cov}(C^{{\rm GW}g}_\ell,C^{{\rm GW}g}_{\ell'})
      &=\delta_{\ell\ell'}\frac{C_\ell^{{\rm GW}{\rm GW}}C_\ell^{gg}+(C_\ell^{{\rm GW}g})^2}{2\ell+1}\\
      &\simeq\delta_{\ell\ell'}\frac{N_\ell C_\ell^{gg}}{2\ell+1},
    \end{align}
    where, in the second equality, we have assumed that noise dominates the GWB signal, and that, as a consequence, $C_\ell^{{\rm GW}g}\ll\sqrt{C_\ell^{gg}N_\ell}$. For simplicity, we assume the same galaxy samples used in this analysis. Since the final uncertainties are dominated by the GWB noise, the impact of shot noise in the galaxy samples, which could be significantly improved with future datasets, can be neglected for this forecast.

    The forecast 95\% upper bounds on $\bOw$ for these future experiments are shown, together with the current constraints for $\alpha=2/3$, in Fig. \ref{fig:bOw_4cast}. Each band in the figure shows the constraints that would be achievable assuming a total observation time from 1 to 10 years. In the most ambitious case, assuming 10 years of observation with the ET-CE network, the combined sensitivity overlaps tantalisingly with the expected signal amplitude. It is therefore conceivable that a future experiment may be able to make a first detection of this signal through cross-correlations. Achieving this, however, will require significant improvements in GW detector technology, and in the design of techniques to exploit detector auto-correlations.

\section{Summary and conclusions}\label{sec:conc}
  We have employed cross-correlations between maps of the SGWB constructed from the publicly available O3 LVK data, and samples of galaxies to place tomographic constraints on the production rate density of gravitational waves over the last $\sim10$ billion years of cosmic history. To do so, we have combined the use of a quadratic minimum variance estimator to measure the cross-correlation optimally weighting the different SGWB map modes given their non-trivial noise properties, with a fast pseudo-$C_\ell$ to measure the galaxy auto-correlation. We find no significant detection of this cross-correlation, and therefore we use our results to place the first upper bound to date on the production rate density of GWs from astrophysical sources $\bOw$, as shown in Fig. \ref{fig:bOw} and Table \ref{tab:zbins}. As expected, given the sensitivity of current observations (see \cite{2005.03001}), our constraints are orders of magnitude above current estimates of the astrophysical signal from binary mergers in galaxies.

  The sensitivity to this noise-dominated measurement will increase rapidly with newer observations, and with fast progress in the design and construction of next-generation gravitational-wave observatories. Although, given current expectations, a detection of the astrophysical anisotropic signal will be extremely challenging, we have shown that future experiments, such as the Einstein Telescope and the Cosmic Explorer may, given long enough observations, reach the sensitivity needed to make this measurement. In this respect, it is worth noting that our current expectation for the anisotropic SGWB from astrophysical sources (i.e. the gray hatched line in Fig. \ref{fig:bOw_4cast}), is subject to large theoretical uncertainties, associated with our limited understanding of the populations of GW sources, and the signal could well be significantly larger (or smaller!) than this expectation. Furthermore, cross-correlations with other cosmological tracers of structure, including CMB lensing \cite{2111.04757}, and the CMB primary anisotropies \cite{2106.02591,2106.03786} may offer alternative avenues to improve the sensitivity of these measurements and potentially uncover unknown signals. Possibly, different GWB mapping techniques such as radiometer mapping~\cite{Mitra:2007mc} which target point-like stochastic sources as opposed to the total angular power spectrum of the background, also regularly employed on LVK data~\cite{O3_SGWB_anisotropic}, may be an alternative avenue to tease out correlated information with galaxy catalogs.
  
  Another promising avenue, covering a completely different frequency range and source type, will be the use of maps constructed from constellations of space-based interferometers \cite{2212.06162}, although the construction of a sufficiently large network probably lies far in the future. Other anisotropic signatures, such as that sourced by a kinematic dipole \cite{2201.10464,2206.02747,2209.11658}, or the large-scale imprint of the Milky Way, may be more easily detectable, and within reach of near-future experiments \cite{2010.00486,2303.15923,2304.06640}.

\section*{Acknowledgements}
  We thank Stefano Camera, Giulia Cusin, Giulio Fabbian, and Kate Storey-Fisher for useful discussions. DA and PGF acknowledge support from STFC and the Beecroft Trust. %
  AIR is supported by the European Union's Horizon 2020 research and innovation programme under the Marie Skłodowska-Curie grant agreement No 101064542, and acknowledges support from the NSF award PHY-1912594. JG is supported by NSF award No. 2207758. %
  We made extensive use of computational resources at the University of Oxford Department of Physics, funded by the John Fell Oxford University Press Research Fund. For the purpose of Open Access, the author has applied a CC BY public copyright licence to any Author Accepted Manuscript version arising from this submission.

\bibliography{main}
\appendix
\onecolumngrid

\end{document}